\documentstyle[multicol,prb,aps,epsf] {revtex}
\begin{document}
\bibliographystyle{prsty}
\draft
\title{Haldane and dimer gap in general double spin-chain models}
\author{Tota Nakamura,$^1$ Satoshi Takada, $^2$ Kiyomi Okamoto,$^3$ 
        and Naoki Kurosawa $^1$}
\address{$^1$ Department of Applied Physics, Tohoku University,
         Sendai, Miyagi 980-77, Japan\\
         $^2$ Institute of Physics, University of Tsukuba,
         Tsukuba, Ibaraki 305, Japan \\
         $^3$ Department of Physics, Tokyo Institute of Technology,
         Oh-Okayama, Meguro, Tokyo 152, Japan
         }
\date{\today}
\maketitle
\begin {abstract}
We have obtained an analytic expression for 
the $k$ dependence of excitation energy gap for an arbitrary 
double $S=1/2$ spin-chain by using the nonlocal unitary transformation 
and the variational method.
It is checked to explain the gap behavior of various systems,
which include the Haldane system and the dimer system in both
extreme limits, and also the ladder model and the Majumdar-Ghosh model.
The string order parameter, the dimer order parameter, 
and the local spin value are also calculated in the ground state.
The ground-state energy exhibits a great stabilization by an
antiferromagnetic bond dimerization,
which might be realized in various new compounds.
We also mention the relation of the convergence to the Haldane state
with the spin-exchange symmetry of the model.
The excited state has one domain wall of a local triplet type
except in the vicinity of the Majumdar-Ghosh point, where
a local triplet is decomposed to two $S=1/2$ free spins moving
among the singlet dimers.
\end  {abstract}

\begin{multicols}{2}

\narrowtext

\section{Introduction}
The low-dimensional quantum systems
with the excitation energy gap have been attracting much interest
both theoretically and experimentally,\cite{dagotto-r96}
though the interest was only from the theoretical side until recently.
The simplest theoretical spin model may be the dimer model that consists of 
independent pairs of $S=1/2$ spins connected by an antiferromagnetic (AF) 
interaction bond.
The ground state is a product of a singlet dimer state on each bond,
and the excitation gap is the singlet-triplet dimer gap.
The Majumdar-Ghosh (M-G) model
\cite{majumdar-g69,shastry-s81}
and the $\Delta$ chain model 
\cite{kubok93,nakamura-k96,sen-swc96}
also realize the perfect singlet dimer ground state, and thus
the gap is intrinsically the dimer gap.
Another well-known model that has a different origin of the gap
is the $S=1$ AF spin chain, so-called the Haldane system.\cite{haldane83}
The ladder model \cite{hida-rev,takada-w92,narushima-nt95,nishiyama-hs95}
and the bond alternation model \cite{hida-rev,hida92,takada92,hida-t92}
interpolate the dimer model and the Haldane system
by changing a strength of interaction bonds as a parameter
from $+\infty$ to $-\infty$.
Therefore, 
these models were investigated mainly to clarify the Haldane system.
Our understandings up to now are that
the dimer state continuously changes to the $S=1$ Haldane state
without any explicit phase transition.
\cite{hida-rev,takada-w92,narushima-nt95,nishiyama-hs95,%
      hida92,takada92,hida-t92}
On the other hand,
the phase transition becomes of the first order in a ladder model with 
both diagonal interactions.\cite{kitatani-o96}

Situation has changed since 
it became possible to synthesize various compounds 
that actually realize the above theoretical models. 
\cite{ramirez94,tanaka-tso96,onoda-n96,takatsu-st97}
For example, magnetic susceptibility measurements on KCuCl$_3$ 
\cite{tanaka-tso96} and on CaV$_2$O$_5$ \cite{onoda-n96}
indicate a spin gap behavior, 
and the experimental data are considered to be explained
through the frustrated double spin-chain model.
\cite{nakamura-o97}
In such an analysis, we need to estimate the
gap as a function of the strength of interaction bonds.
Then the susceptibility can be calculated by using the gap value.
\cite{troyer-tw94}

\begin{figure}[h]
 \epsfxsize = 6.0cm
 \epsffile{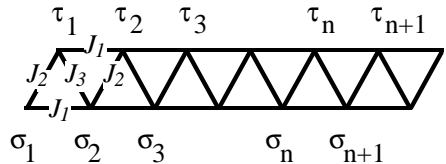}
 \caption {Shape of the general double spin-chain model we treat in this paper.
  \label{fig:lattice}
          }
\end  {figure}

In this paper, we consider the generalized double spin-chain system 
defined by its next-nearest-neighbor interaction, $J_1$, and 
the alternating nearest-neighbor interaction, $J_2$ and $J_3$, as
\begin{eqnarray}
 {\cal H}=\sum_{n=1}^N 
    &J_1&(\mbox{\boldmath $\sigma$}_n \cdot\mbox{\boldmath $\sigma$}_{n+1}
  +     \mbox{\boldmath $\tau  $}_n \cdot\mbox{\boldmath $\tau  $}_{n+1})
\nonumber \\
  + &J_2& \mbox{\boldmath $\sigma$}_n \cdot\mbox{\boldmath $\tau  $}_{n}
  +  J_3  \mbox{\boldmath $\tau  $}_n \cdot\mbox{\boldmath $\sigma$}_{n+1}.
\label{eq:hamiltonian}
\end  {eqnarray}
Here, $N$ is the linear size of the system, 
and $|\mbox{\boldmath $\sigma$}|=|\mbox{\boldmath $\tau$}|=1/2$.
Figure \ref{fig:lattice} shows the depicted lattice.
The system is reduced to the M-G model
with a choice of parameter set $(J_1, J_2, J_3)=(0.5, 1, 1)$,
the isotropic ladder model with $(J_1, J_2, J_3)=(1, 1, 0)$, or $(1, 0, 1)$,
and the $S=1$ AF chain in the limit $J_2\to -\infty$.
The case with $J_1=0$ corresponds to the bond-alternation model in 
a single chain.
Recently, the string order parameter \cite{dennijs-r89}
and the energy for the ground state of this system
have been calculated by the matrix-product method.
\cite{brehmer-mn96}
Here,
we make use of the nonlocal unitary transformation 
\cite{kennedy-t92,takada-k91,nakamura-t97}
and give explicit expressions of
the ground-state energy,
the excitation gap, string order parameter, the dimer order parameter,
local spin value in the ground state, 
and the domain wall spin value in the excited state
for arbitrary $(J_1, J_2, J_3)$.
This transformation is 
an adaptation of the Kennedy-Tasaki transformation \cite{kennedy-t92}
of the $S=1$ system to the double $S=1/2$ spin-chain systems, and is 
known to be powerful when the ground state is either
in the Haldane state or in the state with strong dimer correlation.

In Sec. \ref{sec:groundstate}, we introduce the transformation and the 
variational method employed in this paper.
Then the energy, a local bond-spin value 
and the order parameters for the ground state 
are estimated and compared with the numerical diagonalization results
of the $N=12$ lattice.
Section \ref{sec:excitedstate} describes the excited states,
where we consider two types of the variation.
One is the local-triplet domain wall excitation, which 
we give an explicit form for in the whole phase space.
The other one is what we call the kink-antikink excitation which is
governed by two $S=1/2$ free spins moving among the perfect singlet
dimers.\cite{nakamura-t97,nakamura-t96}
This type is the elementary excitation near the M-G point.

%%%%%%%%%%%%%%%%%%%%%%%%%%%%%%%%%%%%%%%%%%%%%%%%%%%%%%%%%%%%%%%%%%%%%%%%%%%%
\section{Ground state}
\label {sec:groundstate}
%%%%%%%%%%%%%%%%%%%%%%%%%%%%%%%%%%%%%%%%%%%%%%%%%%%%%%%%%%%%%%%%%%%%%%%%%%%%
We first rewrite the Hamiltonian, Eq. (\ref{eq:hamiltonian}), 
with the nonlocal unitary transformation.
\cite{kennedy-t92,takada-k91,nakamura-t97}
The transformation is defined by $U$ in the following.
\begin{eqnarray}
 U&=&\prod_{n=1}^{N} U_n, \\
 U_n&=&P_n^+ + P_n^- \exp[i\pi S_n^x],     \\
 P_n^{\pm}&=&\frac{1}{2}\left(1\pm \exp
                       \left[i\pi\sum_{k=1}^{n-1}S_k^z\right]\right),\\
 \mbox{\boldmath $S$}_n &=& \mbox{\boldmath $\sigma$}_n 
                        + \mbox{\boldmath $\tau  $}_n,
\end  {eqnarray}
where $P_n^+$ ($P_n^-$) is the projection operator onto
states with the even (odd) number of $S_i^z=\pm 1$ for $i\le n-1$.
Then the Hamiltonian (\ref{eq:hamiltonian}) is transformed as,
\begin{eqnarray}
 U^{-1}{\cal H}U=\sum_{n=1}^N 
 &J_1&( - \sigma_n^x\tau  _{n+1}^x - \tau  _n^z\sigma_{n+1}^z
           -4\sigma_n^x\tau  _{n+1}^x   \tau  _n^z\sigma_{n+1}^z )\nonumber \\
+&J_1&( - \tau  _n^x\sigma_{n+1}^x - \sigma_n^z\tau  _{n+1}^z
           -4\tau  _n^x\sigma_{n+1}^x   \sigma_n^z\tau  _{n+1}^z )\nonumber \\
+&J_3&( - \tau  _n^x\tau  _{n+1}^x - \sigma_n^z\sigma_{n+1}^z
           -4\tau  _n^x\tau  _{n+1}^x   \sigma_n^z\sigma_{n+1}^z )\nonumber \\
+&J_2&\mbox{\boldmath $\sigma$}_n \cdot\mbox{\boldmath $\tau$}_{n}.
\end  {eqnarray}
We consider the following variational basis for the ground state of this 
Hamiltonian.
\begin{eqnarray}
 |\Psi_0\rangle &=&\prod_{n=1}^N |n(\alpha, \beta, \gamma, b)\rangle =
          \prod_{n=1}^N( b|T_n\rangle + \sqrt{1-b^2}|S_n\rangle) 
    \label{eq:varbase}               \\
 |S_n\rangle &=&(|\uparrow, \downarrow\rangle 
                -|\downarrow, \uparrow\rangle)/\sqrt{2} \\
 |T_n\rangle &=&\alpha |\uparrow,   \uparrow  \rangle
               +\beta (|\uparrow,   \downarrow\rangle 
                      +|\downarrow, \uparrow  \rangle)/\sqrt{2}
               +\gamma |\downarrow, \downarrow\rangle
\end  {eqnarray}
$|\uparrow, \uparrow\rangle$'s are the states of 
$|\sigma_n^z, \tau_n^z\rangle$.
$b, \alpha, \beta, \gamma$ are the real variational parameters and satisfy 
the normalization condition, $\alpha ^2 + \beta ^2 + \gamma ^2 =1$.
These parameters are supposed to be invariant of $n$,
since we consider the uniform ground state.
In this sense, the present analysis is variational.

A state with $b=0$ is a singlet dimer state on the 
$\mbox{\boldmath $\sigma$}_{n  }$-$\mbox{\boldmath $\tau  $}_n$ bond,
a state with $b=\sqrt{3}/2$ is the other singlet dimer state on the
$\mbox{\boldmath $\sigma$}_{n+1}$-$\mbox{\boldmath $\tau  $}_n$ bond, and
a state with $b=1$ corresponds to the pure VBS state on the
$\mbox{\boldmath $\sigma$}_{n  }$-$\mbox{\boldmath $\tau  $}_n$ bond.
It should be noted that our approximation is not the single-site 
approximation regarding the original Hamiltonian
so that the $\mbox{\boldmath $\sigma$}_{n+1}$-$\mbox{\boldmath $\tau$}_{n}$
dimer can be represented by Eq. (\ref{eq:varbase}) with
$b=\sqrt{3}/2$.

The energy expectation value is calculated as
\begin{eqnarray}
  \langle \Psi_0|\frac{\cal H}{N}| \Psi_0 \rangle
&=& J_2 \left(b^2-\frac{3}{4}\right)
   \nonumber \\
&-& (2J_1+J_3)b^4\left[
          \beta^2(\alpha^2+\gamma^2)+\frac{(\alpha^2-\gamma^2)^2}{4}\right]
   \nonumber \\
&+& (2J_1-J_3)b^2(1-b^2) 
   \nonumber \\
&-& 3J_3 b^3\sqrt{1-b^2}\beta (\alpha^2-\gamma^2)
\end  {eqnarray}
We can easily find this minimum value under the constraint,
$\alpha^2+\beta^2+\gamma^2=1$, by using the Lagrange multiplier.
The energy expectation value $\epsilon_0$ is
%
%\begin{eqnarray}
%\epsilon_0 &=& J_2       \left(b^2-\frac{3}{4}\right) 
%            -  (2J_1+J_3)\frac{b^4}{3}
% \nonumber \\
%           &+& (2J_1-J_3)b^2(1-b^2)
%            -  J_3       \frac{2}{\sqrt{3}}b^3\sqrt{1-b^2},
%\end  {eqnarray}
\begin{equation}
 \epsilon_0 = \left(J_2-\frac{8}{3}b^2J_1\right)
              \left(b^2-\frac{3}{4}\right)
            -\frac{b^2}{3}J_3(b+\sqrt{3(1-b^2)})^2
\end  {equation}
with four possible choices of the parameters $(\alpha, \beta, \gamma)$ as
\begin{eqnarray}
(\alpha, \beta, \gamma)&=&(\pm \sqrt{2/3}, \sqrt{1/3}, 0),\nonumber \\
                        &&(0, -\sqrt{1/3},\pm \sqrt{2/3}),
 \label{eq:abg}
\end  {eqnarray}
and $b$ determined implicitly through
%\begin{eqnarray}
%  J_2&=&(2J_1+J_3)\frac{2b^2}{3} 
%     +(2J_1-J_3)(1-2b^2)
% \nonumber \\
%     &-&J_3 \frac{(3-4b^2)b}{\sqrt{3(1-b^2)}}
%  \label{eq:bdet}
%\end  {eqnarray}
\begin{equation}
 J_2=\left(\frac{4}{3}b^2-1\right)
     \left(4J_1-J_3-\frac{3bJ_3}{\sqrt{3(1-b^2)}}\right)+2J_1
   \label{eq:bdet}
\end  {equation}
or
\begin{equation}
  b=0.
\end  {equation}
The four-fold degeneracy in the choice of $(\alpha, \beta, \gamma)$
corresponds to the degeneracy of the edge states.
\cite{takada92,takada-k91,fath-s93}
A state with $b=0$ is a trivial singlet dimer ground state at $J_2 = \infty$.
The other one, Eq. (\ref{eq:bdet}), represents a nontrivial state that 
will be the ground state for the most of the parameter space.
We solved Eq. (\ref{eq:bdet}) numerically by the bisection method
for arbitrary $(J_1, J_2, J_3)$.

Before going through the details of the following variational results,
let us notice that 
the system possesses the symmetry which exchanges $J_2$ bonds and $J_3$ bonds.
It does not matter if we solve the eigenvalue problem exactly, however, 
the variational results are dependent upon this exchange. 
Therefore, 
we must do the variation on a system
whose $J_2$ and $J_3$ values are exchanged as well as 
on a system with a given parameter $(J_1, J_2, J_3)$,
and must compare both results.
In the present analysis, the ground-state energy is always 
lower if we exchange $J_2$ and $J_3$ in the case of $J_2 > J_3$.

\begin{figure}[ht]
 \epsfxsize = 8.5cm
 \epsffile{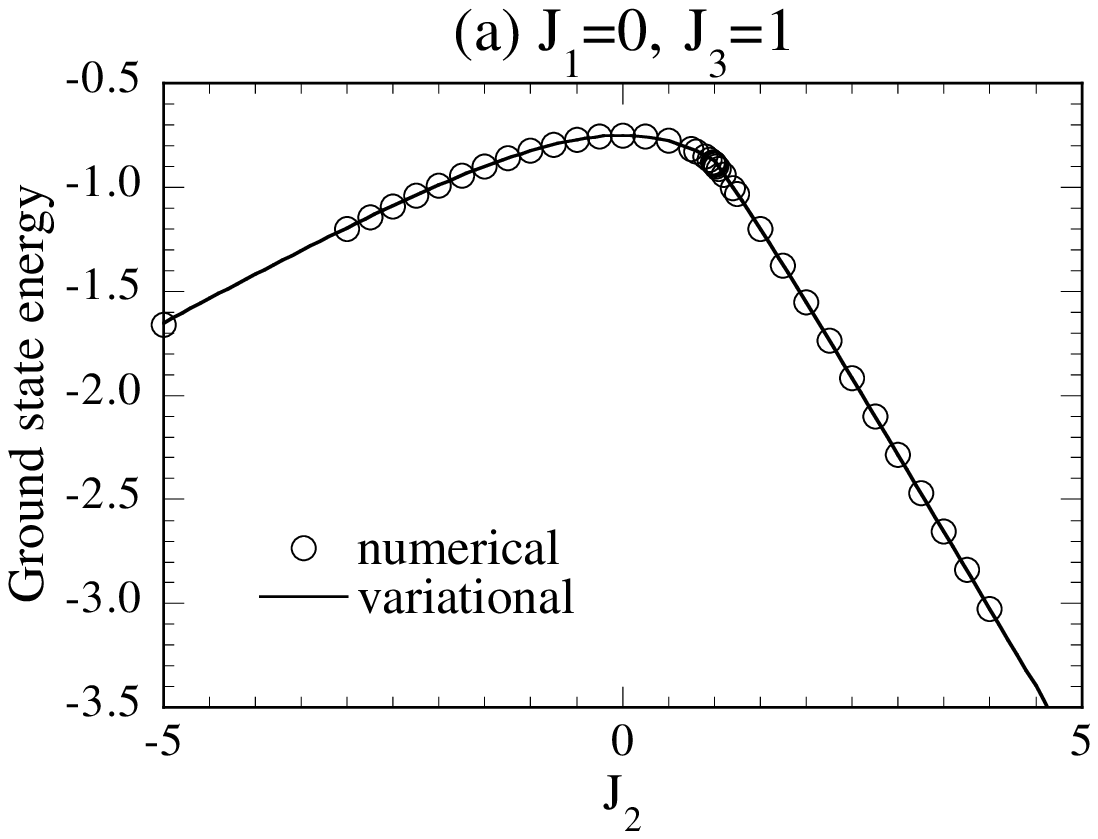}
 \epsfxsize = 8.5cm
 \epsffile{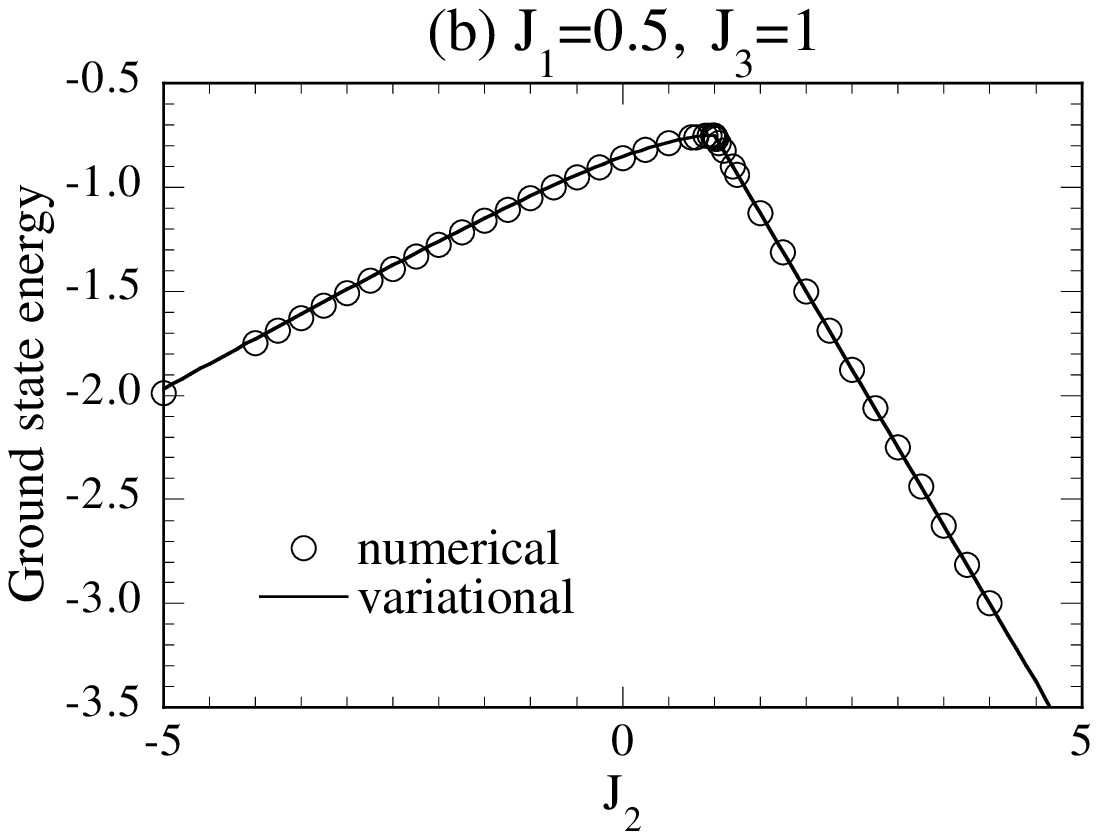}
 \epsfxsize = 8.5cm
 \epsffile{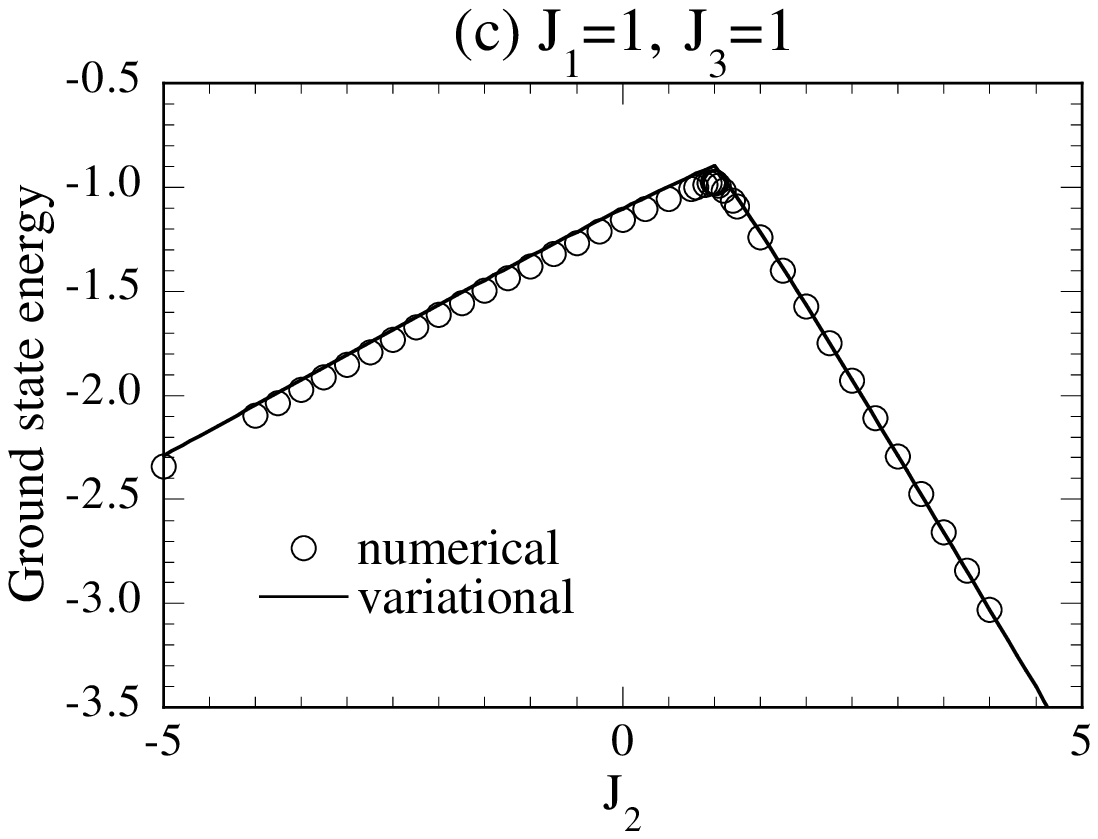}
 \caption {The ground-state energy for
           (a) $J_1=0$,   and $J_3=1$ 
              (bond-alternation model),
           (b) $J_1=0.5$, and $J_3=1$ 
              (includes the M-G model at $J_2=1$), and 
           (c) $J_1=1$, and $J_3=1$,
              (includes the isotropic ladder model at $J_2=0$), respectively.
           Circles denotes the numerical diagonalization results of $N=12$,
           and lines are the variational estimates.
  \label{fig:gsE}
          }
\end  {figure}

We consider three cases which only differ a choice of $J_1$.
$J_3$ is always set equal to $1$, and
$J_2$ is a variable that we move from $+\infty$ to $-\infty$.
The case (a):$J_1=0$ and $J_3=1$ is the bond-alternation model.
It includes a pure dimer model
at $J_2=0$, and the uniform $S=1/2$ AF spin chain at $J_2=1$.
The case (b):$J_1=0.5$ and $J_3=1$ includes the M-G model at $J_2=1$.
The case (c):$J_1=1$ and $J_3=1$ includes the isotropic ladder model 
at $J_2=0$.
In addition, these three cases have a pure dimer model at $J_2=+\infty$ and
the $S=1$ AF chain at $J_2=-\infty$.
Let us call the singlet dimer state on the $J_2$ bonds the $J_2$-singlet, 
and also that on the $J_3$ bonds the $J_3$-singlet hereafter for simplicity.

Figure \ref{fig:gsE} shows the $J_2$ dependence of the ground-state energy,
$\epsilon_0$, compared with the numerical diagonalization results of a system 
with $N=12$ under the periodic boundary conditions.
The variational estimates are done by exchanging $J_2$ and $J_3$ in the 
region $J_2 > J_3$.
The energy agrees with the numerical ones fairly well,
particularly in the dimer region $(J_2 > J_3=1)$.
The difference becomes visible in the Haldane region ($J_2 < J_3=1$)
as $J_1$ increase.
The energy takes maximum at the fully-frustrated point $J_2=J_3=1$ in 
Fig. \ref{fig:gsE} (b) and (c).
As $J_2$ goes away from $J_3=1$, 
the energy decreases because the frustration is relaxed.
It should be noted that this energy stabilization is stronger in the 
dimer region compared with the Haldane region.
Thus, we point out here a possibility that the ground state of the real 
double spin-chain compounds
are also easily stabilized to the dimer ground state by a lattice dimerization
similar to the spin-Peierls system corresponding to the case
Fig. \ref{fig:gsE}(a).

Next, we calculate the local bond-spin value defined by
\begin{eqnarray}
 \langle S(J_2) \rangle  &=&
 \langle \mbox{\boldmath $\sigma$}_n \cdot 
         \mbox{\boldmath $\tau  $}_n 
 + \frac{3}{4}  \rangle =b^2 \\
 \langle S(J_3) \rangle  &=&
 \langle \mbox{\boldmath $\sigma$}_{n+1} \cdot 
         \mbox{\boldmath $\tau  $}_n 
 + \frac{3}{4}  \rangle = -\frac{b^2}{3}(b+\sqrt{3(1-b^2)})^2+\frac{3}{4}.
\end  {eqnarray}
Here, $S(J_i)$ denotes the local spin expectation value along the $J_i$-bond.
Since a local bond-spin value is not a good quantum number,
we have to define it by using a projection operator which selects
local triplet component from the composite spin magnitude.
The bond correlation 
$\mbox{\boldmath $\sigma$}_n \cdot \mbox{\boldmath $\tau  $}_n + 3/4$
serves as this projection operator for $S(J_2)$.
$S(J_i)$ takes zero for the $J_i$-singlet state, while
the other bond-spin takes a value of $3/4$.

\begin{figure}[t]
 \epsfxsize = 8.5cm
 \epsffile{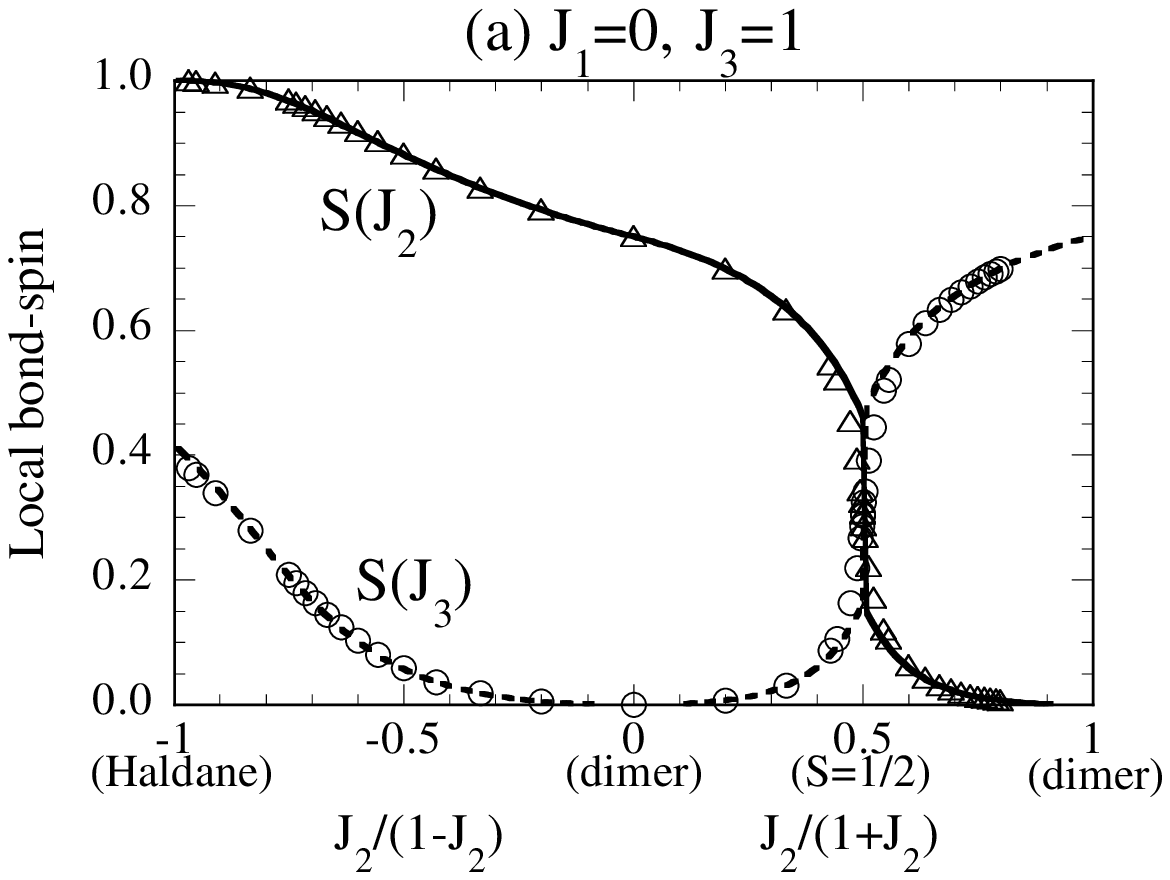}
 \epsfxsize = 8.5cm
 \epsffile{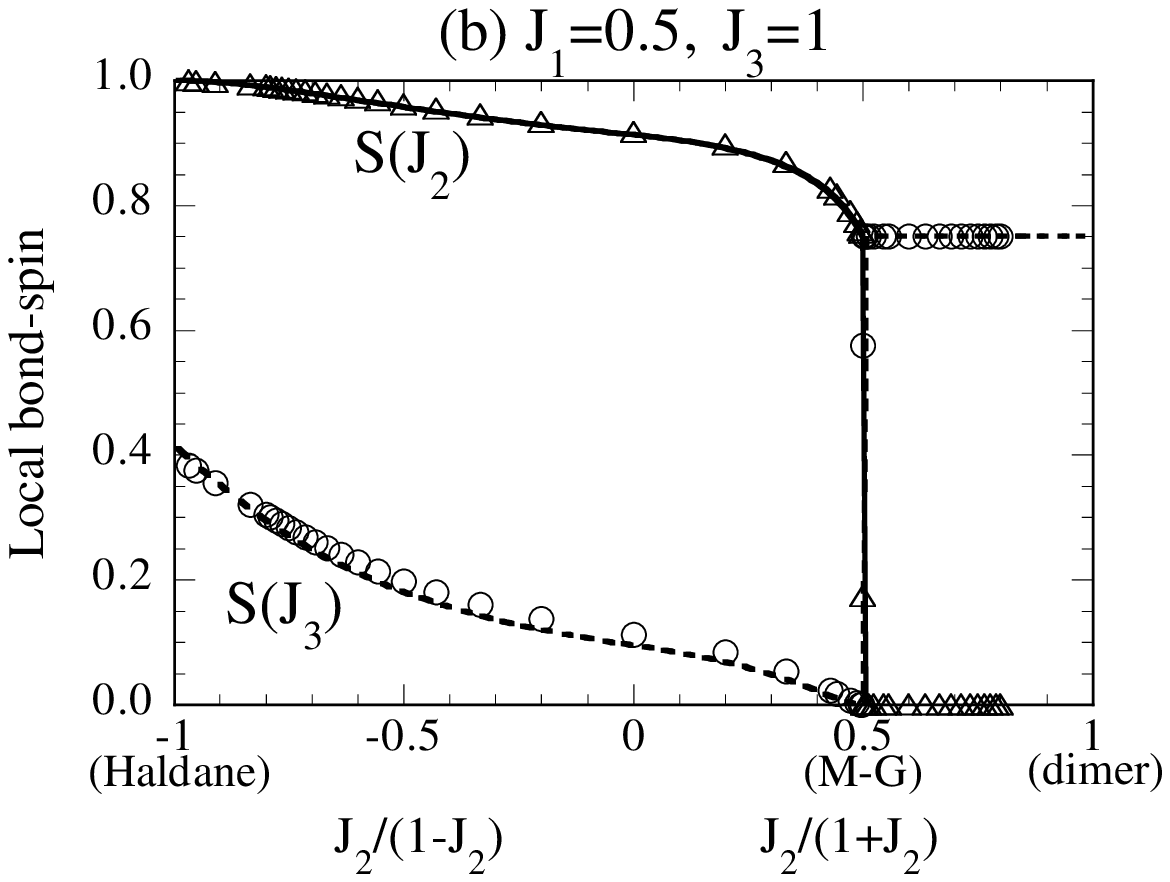}
 \epsfxsize = 8.5cm
 \epsffile{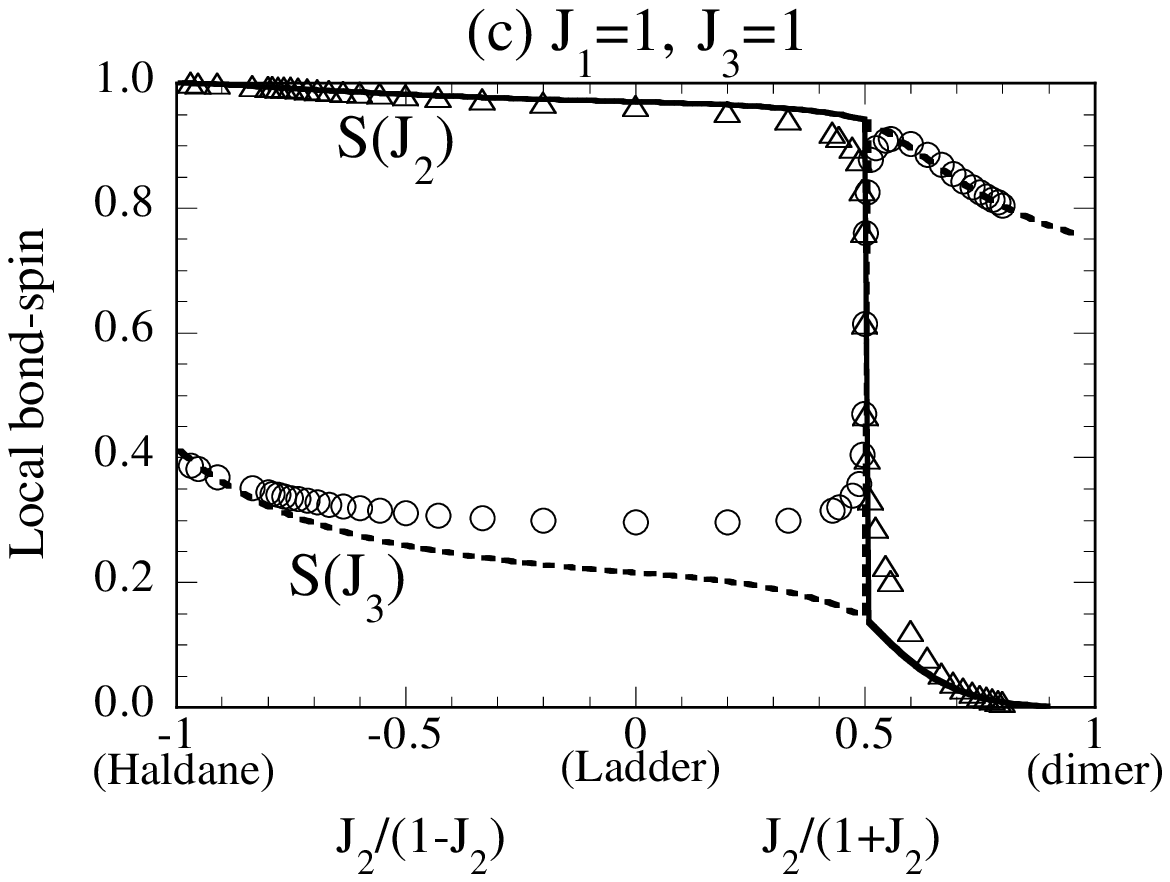}
 \caption {The local bond-spin value on the $J_2$ bond ($S(J_2)$) and 
           that on the $J_3$ bond ($S(J_3)$) plotted
           against $J_2/(1-J_2)$ for $J_2 < 0$, and 
           against $J_2/(1+J_2)$ for $J_2 >0$, where
           (a) $J_1=0$,   and $J_3=1$ 
              (bond-alternation model),
           (b) $J_1=0.5$, and $J_3=1$ 
              (includes the M-G model at $J_2=1$), and 
           (c) $J_1=1$, and $J_3=1$,
              (includes the isotropic ladder model at $J_2=0$), respectively.
           Lines are the variational estimates.
           Triangles and circles are the numerical diagonalization results 
           for $S(J_2)$ and $S(J_3)$
           of a system with 24 spins under the periodic boundary conditions.
 \label{fig:localS}
          }
\end  {figure}

Figure \ref{fig:localS} shows $J_2$ dependence of $S(J_2)$ and $S(J_3)$.
Lines are the variational estimates stated above, and symbols 
are the numerical diagonalization results of $N=12$.

Consistency between the variational estimates and the numerical results
is generally excellent except for the $S(J_3)$ of Fig. \ref{fig:localS}(c).
In Fig. \ref{fig:localS}(a), namely the bond-alternation model,
there are two trivial pure dimer points.
The ground state is the $J_2$-singlet state at $J_2=+\infty$, and 
is the $J_3$-singlet state at $J_2=0$.
These two points are equivalent to each other if we exchange the
$J_2$ bonds and the $J_3$ bonds.
Thus, the local bond-spin values are symmetric 
from the isotropic point at $J_2=1$,
where the model reduces to the uniform $S=1/2$ AF spin chain.
The $J_3$-singlet state continuously changes to the Haldane state 
in the limit of $J_2\to -\infty$, as is visible by 
the $S(J_2)$ converging to 1.
In the model that includes the M-G model, Fig. \ref{fig:localS} (b),
the ground state is exactly the $J_2$-singlet for $J_2>1$,
where $S(J_2)=0$ and $S(J_3)=3/4$.
They show a sudden jump at $J_2=1$, since the $J_3$-singlet is degenerate 
with the $J_2$-singlet at this point.
Situation is rather different in Fig. \ref{fig:localS} (c),
since the $J_3$-singlet never becomes the ground state in this case.
$S(J_3)$ increases from $3/4$ as $J_2$ decrease from $+\infty$ until 
it suddenly decreases at the symmetric point, $J_2=J_3=1$.
$S(J_2)$ now takes nearly the triplet value at $J_2 \sim 1$.
At the isotropic ladder point of $J_2=0$, 
the diagonal spins of the ladder form an almost triplet state 
since $S(J_2)\sim 1$, 
and the rung spins are far from the singlet state, since $S(J_3) \sim 0.3$.

We can also estimate the string order parameter of den Nijs and Rommelse,
$O_{\rm str}$, \cite{dennijs-r89}
and the dimer order parameter, $O_{\rm dim}$
defined by Hida.\cite{hida92,takada92}
They are
\begin{eqnarray}
%O_{\rm dim}(J_2)&=&\lim_{|m-n|\to\infty}-4\langle U^{-1}\sigma_m^z
%      \nonumber \\
%           && \exp [i\pi ( \tau_m^z+ \sum_{k=m+1}^{n-1} S_k^z +\sigma_n^z ) ]
%            \tau_n^z U \rangle                  \nonumber \\
%        &=&\lim_{|m-n|\to\infty} 
%           \langle \exp \left[i\pi \sum_{k=m}^{n} S_k^z \right ]
%           \rangle \nonumber \\
%        &=&\lim_{|m-n|\to\infty}\left(1-\frac{4}{3}b^2\right)^{|m-n|+1}
%         \to \delta_{b=0}    \\
%
 O_{\rm dim}(J_3)&=&\lim_{|m-n|\to\infty}-4\langle U^{-1}\tau_m^z
             \exp  \left[i\pi \sum_{k=m+1}^{n-1} S_k^z \right ]
             \sigma_n^z U \rangle                  \nonumber \\
         &=&\lim_{|m-n|\to\infty} 4\langle \sigma_m^z \sigma_n^z \rangle
          = 4\langle \sigma_m^z \rangle \langle \sigma_n^z \rangle
          = 4\langle \sigma_m^z \rangle ^2  \nonumber \\
            &=&\frac{4b^2}{9}(b+\sqrt{3(1-b^2)})^2,\label{eq:odim} \\
 O_{\rm str}(J_2)&=&\lim_{|m-n|\to\infty} -\langle U^{-1} S_m^z
             \exp      [i\pi \sum_{k=m+1}^{n-1} S_k^z       ]
             S_n^z U\rangle                        \nonumber \\
         &=&\lim_{|m-n|\to\infty} \langle S_m^z S_n^z \rangle
          = \langle S_m^z \rangle \langle S_n^z \rangle
          = \langle S_m^z \rangle ^2 \nonumber \\
            &=&\frac{4}{9}b^4.
   \label{eq:ostr}
%
%O_{\rm str}(J_3)&=&\lim_{|m-n|\to\infty} -\langle U^{-1} 
%          (\tau_m^z+\sigma_{m+1}^z)\nonumber \\
%          &&\exp      \left[i\pi \left(\tau_{m+1}^z +
%             \sum_{k=m+1}^{n-1} S_k^z  +\sigma_{n-1}^z     \right)\right]
%           (\tau_{n-1}^z + \sigma_n^z) U\rangle           \nonumber \\
%        &=&\frac{1}{4}O_{\rm dim}(J_2).
%
\end  {eqnarray}
In addition, $O_{\rm dim}(J_2)=1$ only when $b=0$ and otherwise it vanishes;
$O_{\rm str}(J_3)=O_{\rm dim}(J_2)/4$.

\begin{figure}[t]
 \epsfxsize = 8.5cm
 \epsffile{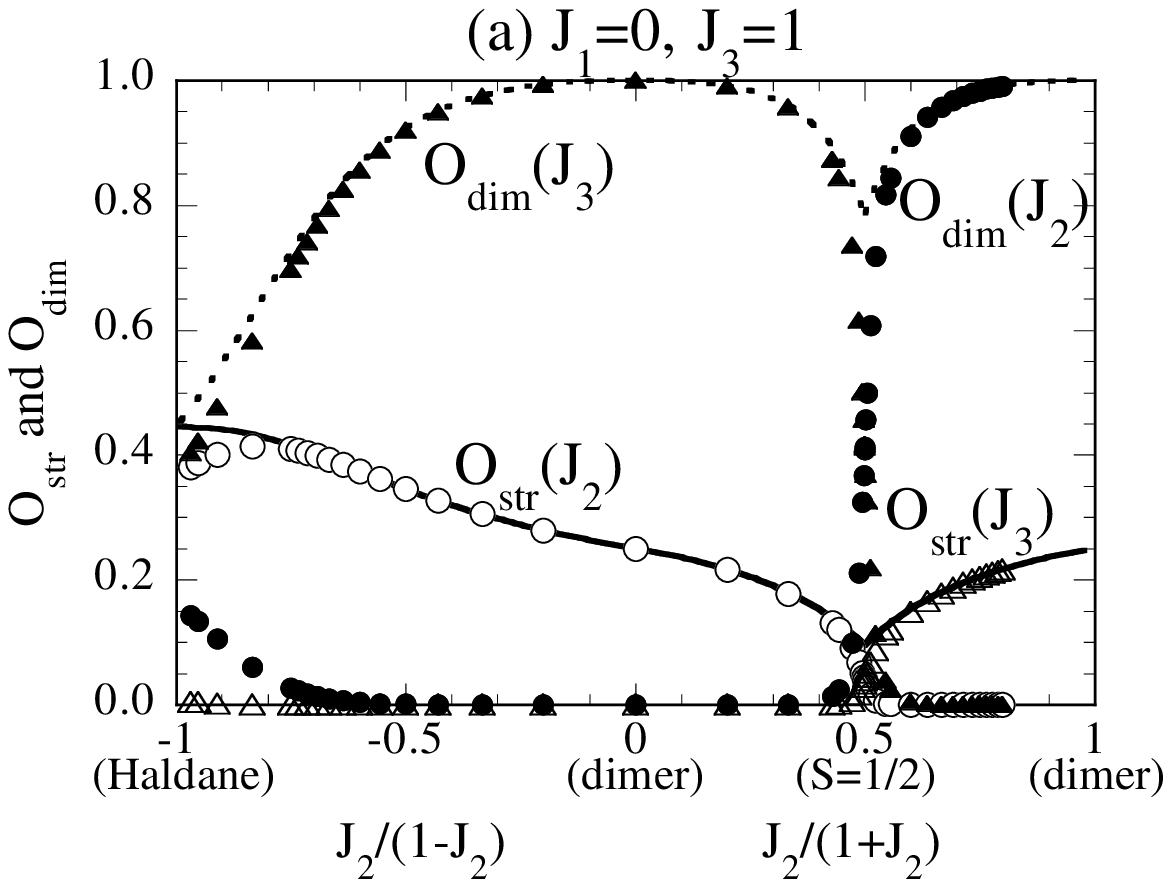}
 \epsfxsize = 8.5cm
 \epsffile{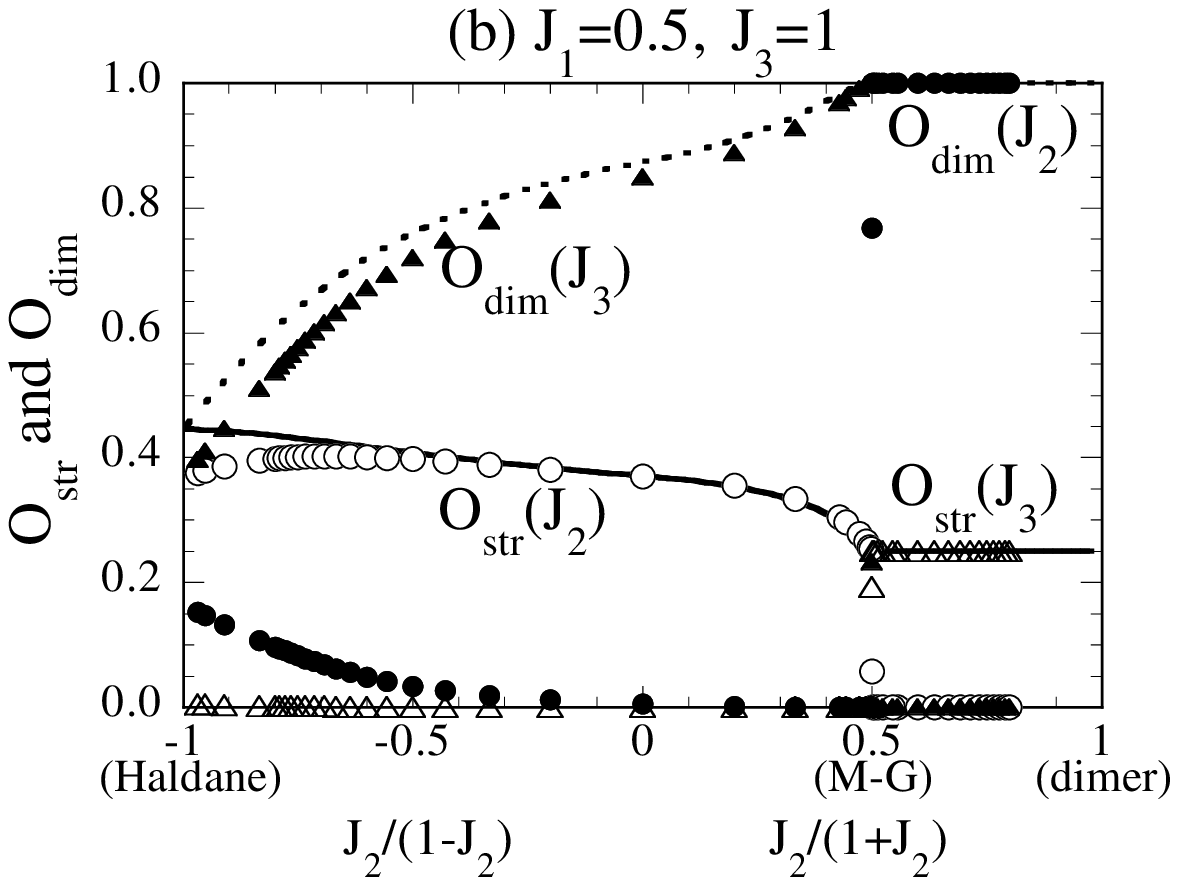}
 \epsfxsize = 8.5cm
 \epsffile{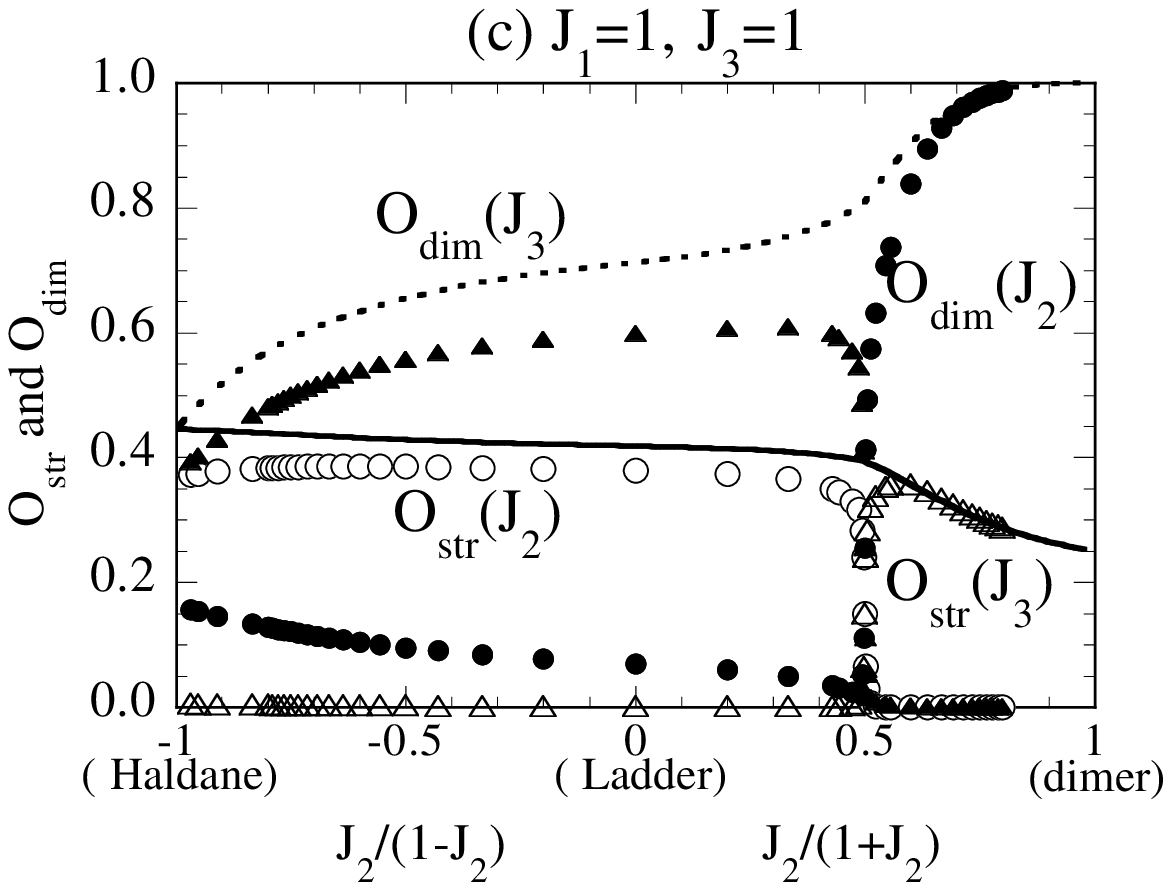}
 \caption {The $O_{\rm str}$ and $O_{\rm dim}$
           on the $J_2$ bonds (circles) and 
           those on the $J_3$ bonds (triangles) are plotted
           against $J_2/(1-J_2)$ for $J_2 < 0$, and 
           against $J_2/(1+J_2)$ for $J_2 >0$, where
           (a) $J_1=0$,   and $J_3=1$ 
              (bond-alternation model),
           (b) $J_1=0.5$, and $J_3=1$ 
              (includes the M-G model at $J_2=1$), and 
           (c) $J_1=1$, and $J_3=1$,
              (includes the isotropic ladder model at $J_2=0$), respectively.
           Solid (broken) lines are the variational estimate 
           for $O_{\rm str}$ ($O_{\rm dim}$).
 \label{fig:str}
          }
\end  {figure}

Figure \ref{fig:str} shows the $J_2$ dependence of 
the dimer and the string order parameter in the 
ground state of three cases mentioned above.
Symbols denote the numerical diagonalization results of $N=12$, 
and lines are the variational estimates.
The order parameters both on the $J_2$ bonds and on the $J_3$ bonds are
plotted in the same figure. 
For example, the $O_{\rm str}(J_2)$ is defined by 
Eq. (\ref{eq:ostr}) when we consider the bond-spin 
$\mbox{\boldmath $S$}_n$ is along the $J_2$-bond, and
the $O_{\rm str}(J_3)$ is defined with $\mbox{\boldmath $S$}_n$
along the $J_3$-bond.
Hence, $O_{\rm str}(J_i)$ denotes the den Nijs-Rommelse string
order regarding the triplet state of each $J_i$-bond.
On the other hand, $O_{\rm dim}(J_i)$ expresses the dimer order
on each $J_i$-bond.
We exchanged $J_2$ and $J_3$ for $J_2 > J_3$ as mentioned before.

Consistency of the variational estimates with the numerical results is
excellent except in the limit of $J_2\to -\infty$.
The variation gives the pure VBS state while the numerical ones converge to 
the correct $S=1$ value.
This is because we used a single-site approximation of the 
transformed Hamiltonian in the variation,
and therefore our estimates always become worse when the correlation length
of the ground state is rather long as is the case in the Haldane state.
Positive values of $O_{\rm dim}(J_2)$ in the region $J_2 < J_3 =1$ are the
finite-size effect, and should vanish in the thermodynamic limit.

In the system of Fig. \ref{fig:str} (a), i.e., the bond-alternation model,
$O_{\rm dim}=1$ at the two pure dimer points.
The variational estimates are particularly good near these two points
because of the short correlation length,
while they become poor near the uniform $S=1/2$ chain point at $J_2=1$ and 
the uniform $S=1$ chain point at $J_2=-\infty$.

In Fig. \ref{fig:str} (b),
the $O_{\rm dim}(J_2)$ is 1 for $J_2 > 1$, 
since the ground state is exactly the $J_2$-singlet.
The $O_{\rm str}(J_2)$ and the $O_{\rm dim}(J_3)$ is zero in this region.
However,
the $O_{\rm str}(J_3)$ takes a finite value of $1/4$. 
This is a natural consequence from the definition 
of the string order parameter.
Thus, it is not adequate to discriminate the Haldane phase 
from the dimer phase by vanishing or non-vanishing 
of this parameter alone.
We should determine the phase by its value
ranging from 1/4 in the dimer state
to $ 0.37$ in the Haldane state.
Therefore,
it is very difficult to draw a phase boundary line in most cases.
As $J_2$ decrease from 1, the $J_3$-singlet remains the ground state
by changing continuously to the
Haldane state in the limit of $J_2\to -\infty$ as is observed in the 
behavior of the $O_{\rm dim}(J_3)$  and the $O_{\rm str}(J_2)$.

The inconsistency between the variational estimates and 
the numerical results in Fig. \ref{fig:str} (b) 
is larger compared with that of (a), 
and becomes distinct in (c).
The difference in Fig. \ref{fig:str} (c) 
is already clear in the vicinity of $J_2=1$, and
remains until $J_2 \to -\infty$.
The variation can explain the behavior of the order parameters only 
qualitatively in this plot.
For example, $O_{\rm str}(J_2)$ is almost invariant of $J_2$ in the
region $J_2 < 1$.

Looking at Figs. \ref{fig:localS} and \ref{fig:str}, we notice
that the convergence to the Haldane state becomes faster
as $J_1$ increases from 0 to 1.
For example, the $O_{\rm str}(J_2)$ of Fig. \ref{fig:str} (c)
takes a value of the Haldane state
even in the isotropic ladder model ($J_2=0$).
This evidence may allow us to consider that 
the ground state of the isotropic ladder model
is more like the Haldane state rather than the dimer state. 

\begin{figure}[t]
 \epsfxsize = 8.5cm
 \epsffile{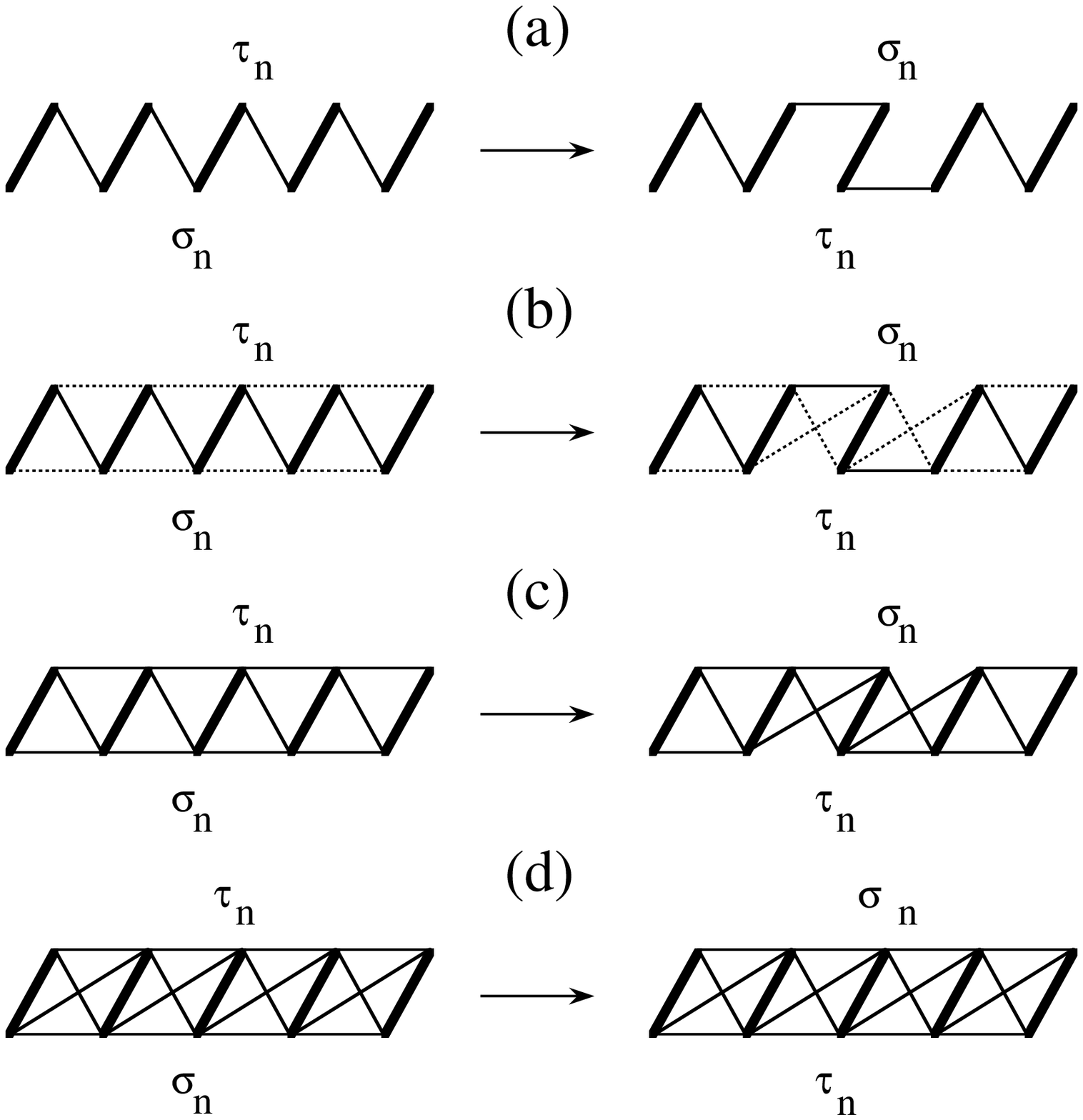}
 \epsfxsize = 8.5cm
 \epsffile{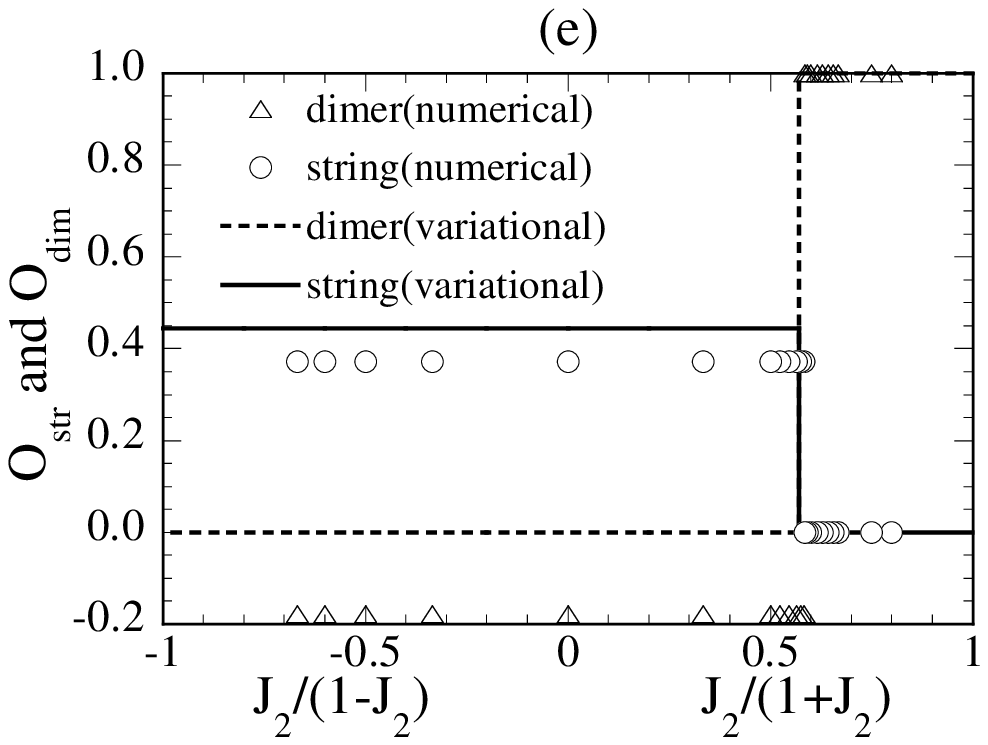}
 \caption {Symmetry of the model with respect to the spin-exchange of
           $\mbox{\boldmath $\sigma$}_n$ and 
           $\mbox{\boldmath $\tau  $}_n$ for
           (a)$J_1=0, J_3=1$,
           (b)$J_1=0.5, J_3=1$,
           (c)$J_1=1, J_3=1$, and
           (d) the ladder model with both diagonal interaction bonds.
           Bold lines denote the $J_2$-bond, thin lines are the bonds with
           their magnitude $1$, and broken lines are those of $0.5$.
           (e)The string and the dimer order parameter of the model (d).
  \label{fig:kuromodel}
          }
\end  {figure}

We relate this tendency to the symmetry of the model with respect to 
the exchange of two spins that couple to form the $S=1$ state 
in the Haldane limit.
In general, this spin-exchange symmetry is necessary 
to realize the Haldane state, since each $S=1$ unit should have this symmetry.
Figure \ref{fig:kuromodel} shows a consequence of an exchange of the spins
$\mbox{\boldmath $\sigma$}_n$ and $\mbox{\boldmath $\tau$}_n$ 
for three models we have considered in this paper.
Bold lines denote the $J_2$ bonds, thin lines are the bonds of their 
magnitude $1$, and broken lines are those of $0.5$.
These models only have this symmetry 
in the limit of $J_2 \to \pm\infty$, and thus 
the Haldane state becomes the exact state only in this limit.
However, the number of the bonds that are invariant before and after this
operation increases as $J_1$ increases from 0 to 1.
In this sense, the case (c) $J_1=1$ and $J_3=1$ is closer to the 
spin-exchange symmetry.

On the other hand,
the ladder model with both diagonal interactions as depicted in 
Fig. \ref{fig:kuromodel}(d) has this symmetry for arbitrary $J_2$ values.
As shown in Fig. \ref{fig:kuromodel}(e),
the first-order transition from the dimer phase to the Haldane 
phase occurs at $J_2=1.40148$.\cite{kitatani-o96}
In this figure, the symbols denote the numerical results of a system 
with 20 spins.
Negative values of the dimer order are the finite-size effect.
Therefore, 
we conclude that the convergence to the Haldane state becomes faster
as the system gains this spin-exchange symmetry.
At the same time, our variational estimates fails faster.

%
%
%

%%%%%%%%%%%%%%%%%%%%%%%%%%%%%%%%%%%%%%%%%%%%%%%%%%%%%%%%%%%%%%%%%%%%
\section{Excited state}
\label {sec:excitedstate}
\begin{figure}[t]
 \epsfxsize = 8.5cm
 \epsffile{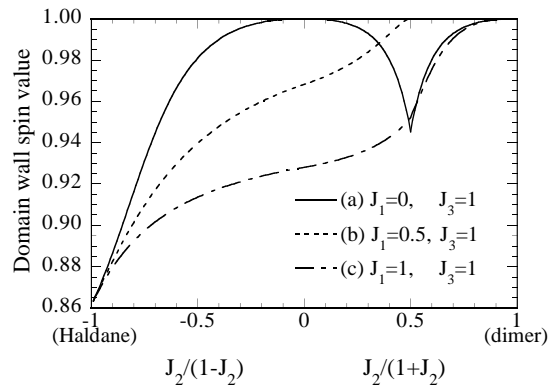}
 \caption {The variational estimates of 
           the spin expectation of a domain wall in the
           excited state are plotted
           against $J_2/(1-J_2)$ for $J_2 < 0$, and 
           against $J_2/(1+J_2)$ for $J_2 >0$, where
           (a) $J_1=0$,   and $J_3=1$ 
              (bond-alternation model),
           (b) $J_1=0.5$, and $J_3=1$ 
              (includes the M-G model at $J_2=1$), and 
           (c) $J_1=1$, and $J_3=1$,
              (includes the isotropic ladder model at $J_2=0$), respectively.
 \label{fig:domainS}
          }
\end  {figure}

The elementary excitation in one dimension is intrinsically 
a state with one domain wall between the degenerate ground states.
Within the present variational scheme, we consider the following 
one domain-wall state under the open boundary conditions.
\begin{eqnarray}
 |\Psi_1\rangle 
                &=& \sum_i C_i 
                 \left (
                \prod_{n=  1}^i| n(\alpha, \beta, \gamma, b)\rangle 
                \prod_{n=i+1}^N| n(\alpha', \beta', \gamma', b)\rangle
                \right )\nonumber \\
                &\equiv& \sum_i C_i \psi_i. 
\end  {eqnarray}
Here, the parameter sets $(\alpha, \beta, \gamma)$ and 
$(\alpha', \beta', \gamma')$ are any two of the four possible choices
given in Eq. (\ref{eq:abg}), and $b$ is determined by Eq. (\ref{eq:bdet}).
For example, we use the set
$(\alpha, \beta, \gamma)=(\sqrt{2/3},\sqrt{1/3},0)$
and $(\alpha', \beta', \gamma')=(-\sqrt{2/3},\sqrt{1/3},0)$.
Of course, the choice does not affect the final results.
A domain wall is located between the $i$th site and the $(i+1)$th site.
This definition of the trial function becomes equivalent to the 
solitonic excitation of F\'ath and S\'olyom \cite{fath-s93}
in the AKLT model.\cite{aklt}
The spin expectation of the domain wall is defined by 
\begin{eqnarray}
&&\langle \psi_i|
\mbox{\boldmath $\tau$}_i\cdot \mbox{\boldmath $\sigma$}_{i+1}+{3}/{4}|
 \psi_i \rangle \nonumber \\
&= &\frac{b^2}{9}(b+\sqrt{3(1-b^2)})^2 + \frac{3}{4}
    =\frac{1}{4}O_{\rm dim}+\frac{3}{4}.
\end  {eqnarray}
Figure \ref{fig:domainS} shows the $J_2$ dependence of 
this value for three different cases discussed in the previous section.
The excitation becomes a local triplet at the domain wall
when the ground state is exactly the singlet dimer state.
As the ground state changes to the Haldane state, the local triplet
smears out and consequently the local spin value at the domain wall
decreases, 
since the total spin of the excited state is always $1$ in this system.
In the Haldane limit, it takes a value of $31/36\sim 0.861$.

The basis relations and the matrix element of the Hamiltonian is calculated
as
\begin{eqnarray}
  \langle\psi_i|\psi_j\rangle &=& \left(1-\frac{4}{3}b^2\right)^{|i-j|} 
                               \equiv (-a)^{|i-j|},\label{eq:basex}\\
  \langle\psi_i|{\cal H}|\psi_j\rangle &=&
     [E_{\rm g}+(|i-j|-1)E_1] \langle\psi_i|\psi_j\rangle \nonumber \\
     &+&\delta _{ij}[E_1+E_2],
\end  {eqnarray}
with
\begin{eqnarray}
 E_{\rm g}&=&\epsilon_0 N  \\
 E_1      &=&-\frac{2b^2}{3}\Bigg[
             2(2b^2-1)J_1 + \frac{3ab}{(b+\sqrt{3(1-b^2)})^3}J_3 \nonumber \\
          &+&  \frac{1}{2}\left[(b+\sqrt{3(1-b^2)})^2-1\right]J_3 \Bigg ] \\
 E_2      &=& \frac{4b^2}{9}\Bigg[
              6aJ_1 + (b+\sqrt{3(1-b^2)})^2J_3 \Bigg].
\end  {eqnarray}
The variation
$
 {\langle \Psi_1|{\cal H}|\Psi_1\rangle}
/{\langle \Psi_1|         \Psi_1\rangle}
$
can be calculated by the Fourier transformation,
since the denominator $\langle \Psi_1|\Psi_1\rangle$ is diagonalized
in the thermodynamic limit.
Namely, by $|\phi_k\rangle = \sum_n \exp[ikn]|\psi_n\rangle$.
Then the energy gap is obtained with respect to the wave number 
of the domain wall $k$ as,
\begin{eqnarray}
  E_{\rm ex}(k)&=&
         \frac{\langle \phi_k|{\cal H}|\phi_k\rangle}
              {\langle \phi_k|         \phi_k\rangle}-E_{\rm g}
          \nonumber \\
          &=&
         -E_1 \left( 
         1+\frac{2a}{1-a^2}
           \frac{(1+a^2)\cos k + 2a}{1+2a\cos k + a^2} \right)\nonumber \\
         &+& (E_1+E_2)\frac{1+2a\cos k + a^2}{1-a^2},
 %       &+&E_2 \frac{1+2a\cos k + a^2}{1-a^2},
  \label{eq:gaplocal1}
\end  {eqnarray}
where $a$ is defined by Eq. (\ref{eq:basex}).

\begin{figure}[t]
 \epsfxsize=8.5cm
 \epsffile{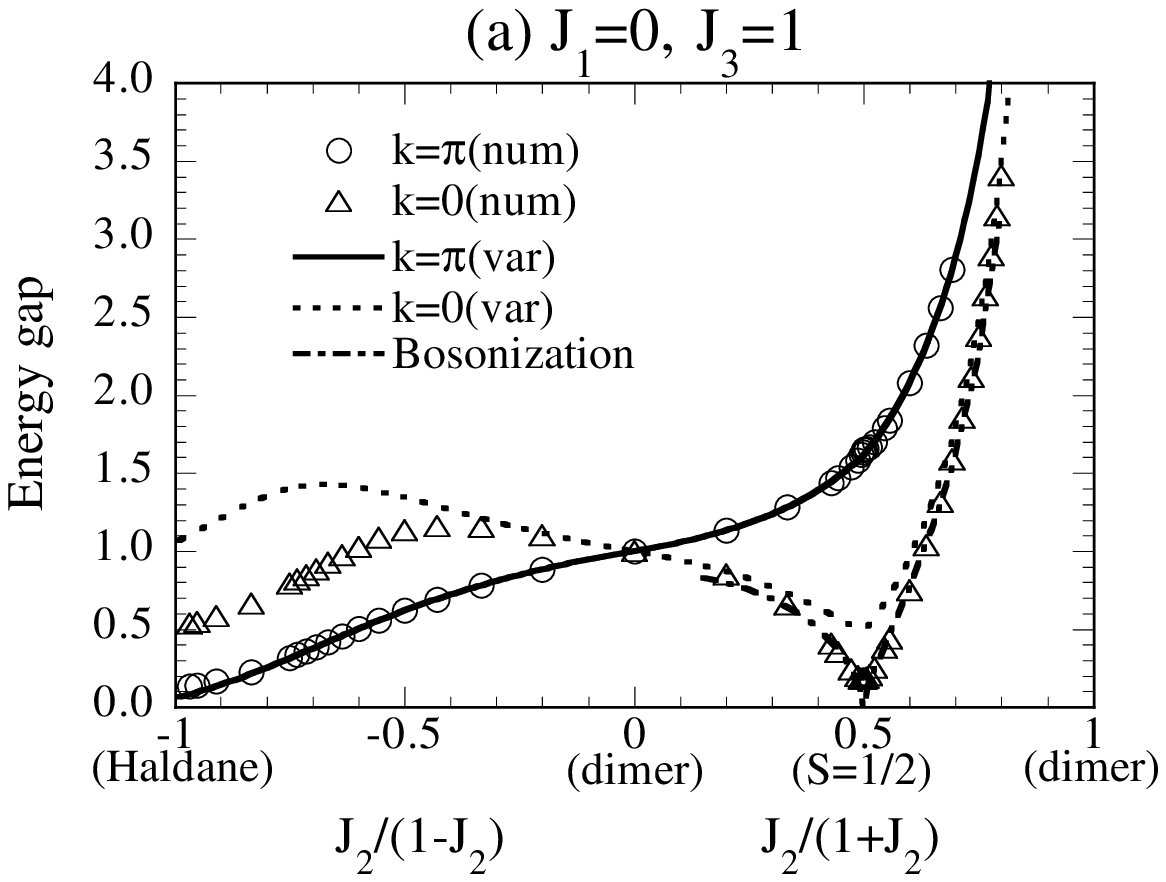}
 \epsfxsize=8.5cm
 \epsffile{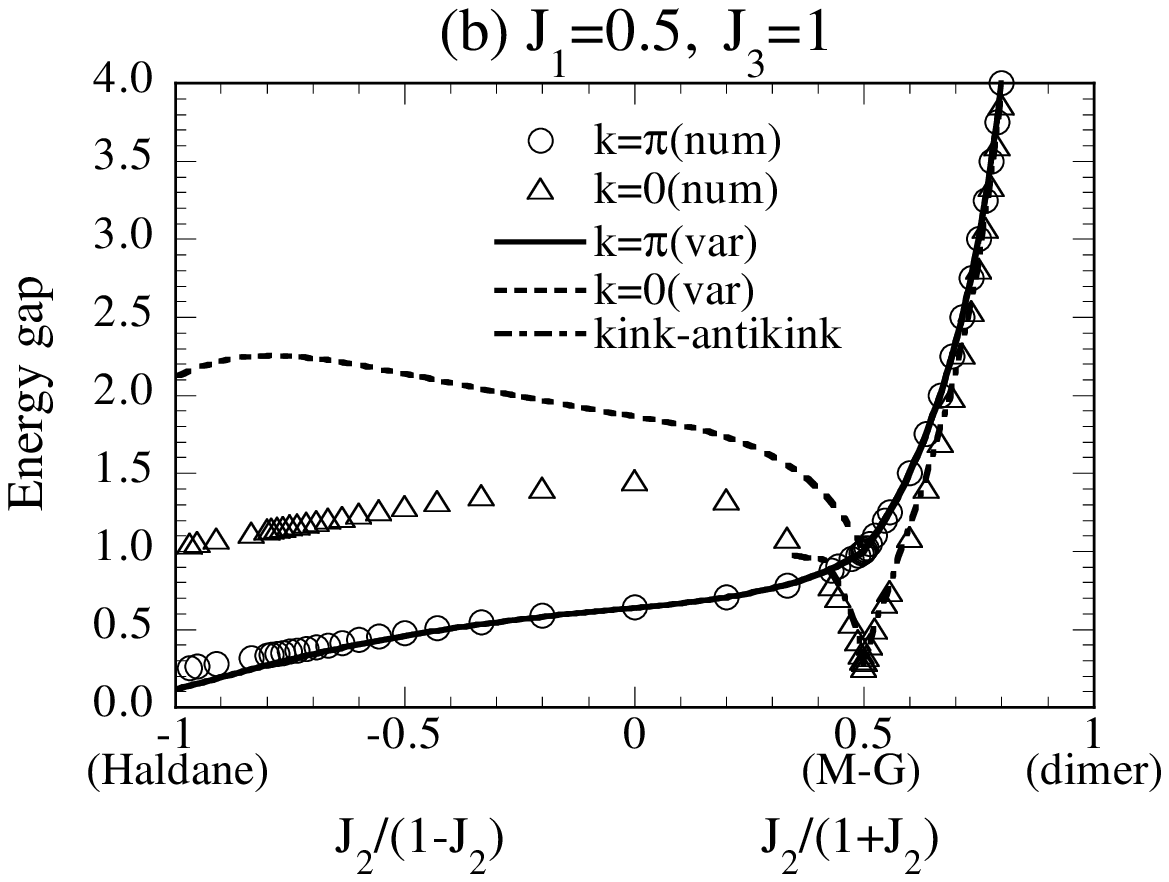}
 \epsfxsize=8.5cm
 \epsffile{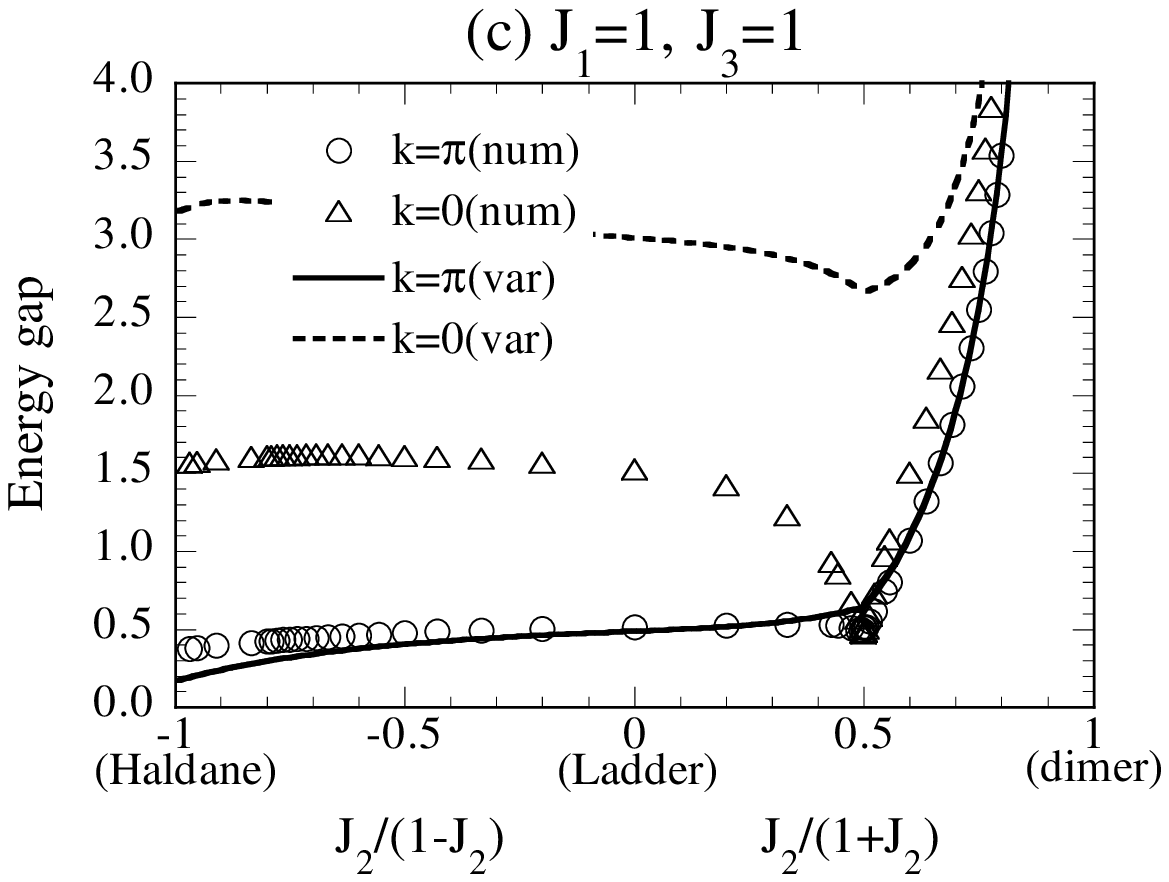}
 \caption{$J_2$ dependence of the energy gap obtained by 
          the numerical diagonalization (symbols), and
          the variation (lines) plotted 
           against $J_2/(1-J_2)$ for $J_2 < 0$, and 
           against $J_2/(1+J_2)$ for $J_2 >0$, where
           (a) $J_1=0$,   and $J_3=1$ 
              (bond-alternation model),
           (b) $J_1=0.5$, and $J_3=1$ 
              (includes the M-G model at $J_2=1$), and 
           (c) $J_1=1$, and $J_3=1$,
              (includes the isotropic ladder model at $J_2=0$), respectively.
          We also plot the estimate by 
          the kink-antikink variation (broken dash line) in (b).
 \label{fig:gapall}
         }
\end  {figure}

We plot this estimate with $k=0$ and $k=\pi$ in Fig. \ref{fig:gapall} 
and compare with the numerical results.

In the vicinity of the M-G point, 
$(J_1, J_2, J_3)=(0.5, 1, 1)$,
the lowest excitation is a kink-antikink state, which
consists of $(N-1)$ singlet dimer pairs 
and two free $S=1/2$ spins.\cite{shastry-s81}
We call these free spins a kink and an antikink.
In such a case, we try another variation.
These two free spins are mobile in general and thus 
we have to consider the following 
trial function,
\begin{equation}
  |\psi_{ij}\rangle = 
                \prod_{n=1}^{i-1}| n(0)\rangle 
                \prod_{n=i}^{j-1}| n(\alpha, \beta, \gamma, b)\rangle
                \prod_{n=j}^N    | n(0)\rangle,
\end  {equation}
under the periodic boundary conditions.
Here, $|n(0)\rangle$ stands for the singlet dimer state on the 
$\mbox{\boldmath $\sigma$}_n$-$\mbox{\boldmath $\tau$}_n$ bond with $b=0$.
$|n(\alpha, \beta, \gamma, b)\rangle$ becomes another singlet dimer state
on the $\mbox{\boldmath $\sigma$}_{n+1}$-$\mbox{\boldmath $\tau$}_n$ bond
when $b=\sqrt{3}/2$ at the M-G point.
Therefore a kink or an antikink should exist at the domain wall.

In the $\Delta$ chain, a kink becomes localized and thus the problem can
be reduced to a one-body problem of a moving antikink.
We can solve it exactly and the wave function is given by the Airy function.
\cite{nakamura-t97,nakamura-t96}
Here, we solved this variation only 
numerically for a finite system of $N=30$. 
The results are also plotted in Fig. \ref{fig:gapall}(b) by 
a broken dash line.

The relative motion of this two-body problem is same as the 
one-body problem of the $\Delta$ chain.
That is,
we obtain the same equation
if we set one free spin at the origin and rewrite the variational problem 
with respect to the one body-problem of the other free spin.
This is also verified by that 
the wave function obtained numerically is well 
fitted by the Airy function, and that the gap enhancement,
$E_{\rm gap}(J_2)-E_{\rm gap}(J_2=1)$, is almost equal to that of the 
$\Delta$ chain; it obeys a power law with exponent $2/3$.
\cite{nakamura-t97,nakamura-t96,chitra95}
Of course, the gap itself cannot be obtained correctly only 
by the relative motion but together with the motion of the center of mass.

Figure \ref{fig:gapall} shows the $J_2$ dependence of the energy gap 
estimated above
for the cases with (a) $J_1=0, J_3=1$, 
(b) $J_1=0.5, J_3=1$ and (c) $J_1=1, J_3=1$.
The local triplet excitation with $k=\pi$ is depicted by solid lines,
that with $k=0$ is by broken lines, and
the kink-antikink excitation is by a broken dash line in 
Fig. \ref{fig:gapall}(b).
The lowest gap in the $k=\pi$ sector and that in the $k=0$ sector
calculated by the numerical diagonalization of an $N=12$ lattice
are depicted by circles and by triangles, respectively.
The variational estimates with $k=\pi$ are quite excellent for all the plots.
We consider this is because the local approximation is usually good
for the wave number $\pi$, which changes the phase of the wave function 
by only one lattice spacing.
Only a difference is that the local triplet variation converges to
the VBS value in the $J_2\to -\infty$ limit, while the numerical one 
converges to the Haldane value.
When the ground state is exact singlet dimer state 
($J_2 > J_3=1$ in Fig. \ref{fig:gapall}(b)),
our variational excitation is a local triplet state and thus 
the gap value is equal to $J_2$ with no dispersion.
On the other hand,
the estimates with $k=0$ only explain the gap behavior qualitatively, and
become worse with an increase of $J_1$.
They are only valid near the pure dimer points in Fig. \ref{fig:gapall}(a).
In the vicinity of the gapless point of $J_2=J_3=1$ (S=1/2 chain), 
our variational scheme breaks down, and thus
the gap estimate should be done by another method. 
For example, the bosonization technique 
\cite{nakano-f80,okamoto96,okamoto-n97}
gives 
\begin{equation}
 \frac{1}{2}\left (\frac{18}{\pi}\right)^{\frac{1}{3}}
 \frac{|2x-1|^{\frac{2}{3}}}
      {1-x},
\end  {equation}
with $x=J_2/(1+J_2)$,
and agrees with our numerical results very well.
This is plotted by a broken dash line in Fig. \ref{fig:gapall}(a).
The variational estimate of the kink-antikink type is also consistent 
with the numerical results.
We can consider that it is valid as long as the singlet dimer state is 
exactly or approximately the ground state.

\section{summary and discussion}
\label{sec:sum}
We have investigated the general $S=1/2$ double spin chain
($J_1$-$J_2$-$J_3$ model) by means of the nonlocal unitary transformation 
and the variation.
The model includes the dimer model and the Haldane system in its 
both extremes.
A ground-state change occurs at the symmetric point, $J_2=J_3=1$,
as is visible in the string and the dimer order parameter.
The ground state of $J_2 < J_3 $ continuously changes to the Haldane state 
in the $J_2\to -\infty$ limit, 
and the other ground state of $J_2 > J_3$ becomes the dimer state
in the $J_2\to  \infty$ limit.
Two states are degenerate at the symmetric point.

We relate the convergence to the Haldane state with the symmetry of 
the exchange of spins that will couple to the $S=1$ state in the 
Haldane limit.
As the system gains amount of this symmetry,
the convergence of the ground state to the Haldane state becomes faster,
or in other words, the transition becomes close to the first-order.
In the case of the system with full symmetry, the 
transition is strictly of the first-order.\cite{kitatani-o96}

The excited state is formulated by a domain wall between 
two of the four-fold degenerate ground states.
In the dimer region, this domain wall is equivalent to a local triplet.
We obtained an explicit form for the dispersion relation of the gap
for arbitrary $J_1$, $J_2$, and $J_3$.
That is, we first calculate the value of $b$ by Eq. (\ref{eq:bdet})
for a given $(J_1, J_2, J_3)$.
Then, the dispersion is given by Eq. (\ref{eq:gaplocal1}) with 
$a=4/3b^2-1$.
We confirmed that our variational estimate is generally good in the 
dimer region, $J_2 > J_3$, and is especially excellent for
the excited state with $k=\pi$.
The lowest excitation near the M-G point is 
well-explained by a kink-antikink state.
\cite{shastry-s81,nakamura-t97,nakamura-t96}

Our variation employs the single-site approximation
in the transformed system, which of course
becomes worse when the ground state has rather long correlation length,
e.g. like in the Haldane state.
Therefore, our estimate fails faster with
the convergence to the Haldane system.
We must go beyond the single-site approximation for the sake of quantitative
agreements in this region.

Finally, we point out a possibility that the ground state of a
real compound is stabilized by a lattice dimerization to the dimer state,
since its energy stabilization is quite significant as we observed in 
Fig. \ref{fig:gsE}.
In fact, the susceptibility of KCuCl$_3$ is quantitatively explained 
by a single dimer model with $J=48.8$K.\cite{nakamura-o97}
The situation is similar in the case of CaV$_2$O$_5$.\cite{onoda-n96}

\acknowledgments
%\ack

Authors 
would like to thank M. Onoda and H. Tanaka for valuable discussions.
They also 
acknowledge thanks to H. Nishimori for his 
diagonalization package, Titpack Ver. 2.
Computations were done partly
on Facom VPP500 at the ISSP, University of Tokyo.

%\section*{References}

\end{multicols}
\end{document}